\documentclass[preprint
]{revtex4-1}
\usepackage[dvipdfmx]{graphicx}
\usepackage{bm}
\usepackage{color}
\begin{document}
\title{%
Photoinduced enhancement of excitonic order in the two-orbital Hubbard model}
\date{\today}

\author{Yasuhiro Tanaka}
\email{yasuhiro@phys.chuo-u.ac.jp}
\author{Manabu Daira}
\author{Kenji Yonemitsu}
\affiliation{Department of Physics, Chuo University, Bunkyo, Tokyo 112-8551, Japan}

\begin{abstract}
Photoinduced dynamics in an excitonic insulator is studied theoretically by using a two-orbital Hubbard 
model on the square lattice where the excitonic phase in the ground state is characterized by the BCS-BEC 
crossover as a function of the interorbital Coulomb interaction. 
We consider the case where the order has a wave vector ${\bm Q}=(0,0)$ and photoexcitation is 
introduced by a dipole transition. Within the mean-field approximation, we show that the excitonic 
order can be enhanced by the photoexcitation when the system is initially in the BEC regime of the excitonic 
phase, whereas it is reduced if the system is initially in the BCS regime. The origin of this 
difference is discussed from behaviors of momentum distribution functions and momentum-dependent excitonic pair 
condensation. In particular, we show that the phases of the excitonic pair condensation have an important role 
in determining whether the excitonic order is enhanced or not. 
\end{abstract}

\maketitle

\section{Introduction}
Ultrafast control of electronic properties in materials by light irradiation has become a 
fascinating subject in condensed matter physics. 
In particular, recent experimental findings such as photoinduced localization of 
charges in metals\cite{Ishikawa_NatComm14,Kawakami_PRB17} and appearance of novel 
transient orders by photoexcitation\cite{Onda_PRL08,Fausti_Sci11,Singer_PRL16,Mor_PRL17} 
have attracted much attention. These phenomena are in sharp contrast to conventional 
photoinduced phase transitions typified by insulator-to-metal transitions in 
correlated electron systems\cite{Yonemitsu_PR08} where photoexcitation usually melts 
electronic orders. 

Excitonic insulators (EIs) were proposed to appear in a semimetal 
with a small band overlap\cite{Mott_PM61} or in a semiconductor with a small band 
gap\cite{Knox_SSP63}. This state arises from macroscopic condensation of bound electron-hole 
pairs, excitons, that are mediated by the Coulomb interaction. 
Although theories of EIs have been developed since 1960s\cite{Jerome_PR67,Kunes_JPCM15,Halperin_RMP68}, 
its experimental identification 
is a challenging task. In fact, a few materials such as 
1$T$-TiSe$_2$\cite{Salvo_PRB14,Kidd_PRL02,Cercellier_PRL07,Monney_PRB09,Monney_PRL11,Pillo_PRB00} 
and Ta$_2$NiSe$_5$\cite{Wakisaka_PRL09,Kaneko_PRB13,Seki_PRB14,Lu_NatComm17} have 
been known as candidates for EIs. In this regard, a search for novel photoinduced phenomena 
that are peculiar to EIs is of great interest. 
Among the candidate materials, 1$T$-TiSe$_2$ is a semimetallic material that 
exhibits a charge-density-wave (CDW) state at low temperatures. 
Through studies of photoinduced melting dynamics of the CDW, 
a signature of a possible excitonic order in the CDW, i.e., an 
excitonic CDW has been argued from a nonequilibrium point of 
view\cite{Rohwer_NAT11,Vorobeva_PRL11,Porer_NatMat14,Monney_PRB16,Mathias_NatComm16}. 
Ta$_2$NiSe$_5$ is a semiconductor with a small direct gap above $T_C=326$K where a 
semiconductor-to-insulator transition occurs. It has been shown that various experimental 
results are consistent with the realization of an excitonic order for 
$T<T_C$\cite{Lu_NatComm17}. Quite recently, photoinduced enhancement of the gap associated with the 
excitonic order has been reported for this material\cite{Mor_PRL17}. 

Theoretically, most studies on EIs have been concerned with their equilibrium 
properties\cite{Kaneko_PRB13,Phan_PRB10,Zocher_PRB11,Seki_PRB11,Zenker_PRB12,Kaneko_PRB12,Seki_PRB14,Watanabe_JPCS15,Nasu_PRB16,Tatsuno_JPSJ16}. 
For instance, the BCS-BEC crossover that is an important concept for characterizing the nature 
of the pair condensation has been examined in EIs\cite{Seki_PRB11,Phan_PRB10,Watanabe_JPCS15,Kaneko_PRB12}. 
In nonequilibrium conditions, photoinduced melting dynamics of the excitonic order 
which is accompanied by a CDW has 
been investigated in semimetallic systems\cite{Golez_PRB16}. 
For a direct gap semiconductor like Ta$_2$NiSe$_5$, 
Murakami {\it et al.} have shown that photoexcitation enhances the gap by 
electron-phonon couplings\cite{Murakami_arX17}. In spite of these studies, our 
understandings of photoinduced properties of EIs are still limited. 
For example, a possibility for the gap enhancement in purely electronic models has not been 
elucidated so far. A relevance of the BCS-BEC crossover to photoinduced states has not been 
fully explored yet. 

In this paper, we investigate photoinduced dynamics of EIs using a two-orbital Hubbard model on the square 
lattice, where we consider electric dipole transitions caused by photoexcitation. 
By computing momentum distribution functions and excitonic pair condensation 
in momentum space, we show that when the system initially possesses a BEC-type (BCS-type) excitonic 
order, its dynamics is essentially described by a real (momentum) space picture. 
In this sense, the photoinduced dynamics is strongly affected by where the system is located 
in the BCS-BEC crossover in thermal equilibrium. 
For the BEC-type order, an enhancement of the gap is realized, whereas the gap is reduced 
for the BCS-type order. We explain this difference by analyzing time evolution of the phases 
of the excitonic pair condensation. 

\section{Model and Method}
We consider a two-orbital Hubbard model on the square lattice defined by the following 
Hamiltonian, 
\begin{eqnarray}
\hat{H}&=&t_{c}\sum_{\langle ij\rangle \sigma}(c^{\dagger}_{i\sigma}c_{j\sigma}+h.c.)+\mu_C\sum_{i\sigma}n^{c}_{i\sigma}
+t_{f}\sum_{\langle ij\rangle \sigma}(f^{\dagger}_{i\sigma}f_{j\sigma}+h.c.)\nonumber \\
&+&U\sum_{i}n^{c}_{i\uparrow}n^{c}_{i\downarrow}+U\sum_{i}n^{f}_{i\uparrow}n^{f}_{i\downarrow}+U^{\prime}\sum_{i}n^{c}_{i}n^{f}_{i},
\label{eq:ham}
\end{eqnarray}
where $\alpha^{\dagger}_{i\sigma}$ ($\alpha=c, f$) is the creation operator for an electron 
with spin $\sigma$ ($=\uparrow, \downarrow)$ at the $i$-th site on the $\alpha$ orbital. 
We define $n^{\alpha}_{i\sigma}=\alpha^{\dagger}_{i\sigma}\alpha_{i\sigma}$ and 
$n^{\alpha}_i=n^{\alpha}_{i\uparrow}+n^{\alpha}_{i\downarrow}$. The parameter $t_{\alpha}$ is the 
transfer integral for 
electrons on the $\alpha$ orbital and $\langle ij\rangle$ denotes a pair of nearest-neighbor sites. 
In this paper, we set $t_f=1$ and $t_c=-1$ and choose the former as the unit of energy. 
The quantity $\mu_C(>0)$ is a parameter that controls the overlap between the $c$ and $f$ bands. 
The intraorbital and 
interorbital Coulomb interactions are denoted by $U$ and $U^{\prime}$, respectively. 
We fix the electron density per site at $n=2$. 

For the photoexcitation, we introduce a time $(\tau)$-dependent term that describes electric 
dipole allowed transitions\cite{Golez_PRB16,Murakami_arX17} as 
\begin{equation}
\hat{H}_D(\tau)=F(\tau)\sum_{i\sigma}(c^{\dagger}_{i\sigma}f_{i\sigma}+h.c.),
\label{eq:ham_D}
\end{equation}
where $F(\tau)=F_0\sin(\omega \tau)\theta (\tau)\theta (T_{\rm irr}-\tau)$ with $\theta (\tau)$ being the 
Heaviside step function. The pulse width is denoted by $T_{\rm irr}=2\pi N_{\rm ext}/\omega$ and we use 
single cycle pulses ($N_{\rm ext}=1$) throughout the paper. 

We apply the Hartree-Fock (HF) approximation to Eq. (\ref{eq:ham}). For $U=U^{\prime}=0$, the band 
structure corresponds to a direct gap semiconductor for $\mu_C>8$. For $\mu_C<8$, the Fermi surface 
has electron and hole pockets that coincide with each other owing to the particle-hole symmetry. 
The system has an instability toward excitonic condensation with a wave vector ${\bm Q}=(0,0)$ 
so that we define the excitonic order parameter as 
$\Delta_0=\langle c^{\dagger}_{i\sigma}f_{i\sigma}\rangle$, which is independent of $i$. 
We assume that $\Delta_0$ does not depend on $\sigma$. 
Similarly, we write $\langle n^{\alpha}_{i\sigma}\rangle=n_{\alpha}/2$ where $n_{\alpha}$ is the electron 
density on the $\alpha$ orbital per site. Since $n=2$, we have $n_f=2-n_c$. 
The total Hamiltonian within the HF approximation $\hat{H}^{\rm HF}_{\rm tot}(\tau)=\hat{H}^{\rm HF}+\hat{H}_D(\tau)$ 
is written in momentum representation as 
\begin{eqnarray}
\hat{H}^{\rm HF}_{\rm tot}(\tau)&=&\sum_{k\sigma}\hat{H}_{k\sigma}(\tau),\\
\hat{H}_{k\sigma}(\tau)&=&\Psi^{\dagger}_{k\sigma}h_{k}(\tau)\Psi_{k\sigma},
\label{eq:ham_t}
\end{eqnarray}
where $\Psi^{\dagger}_{k\sigma}=(c^{\dagger}_{k\sigma}, f^{\dagger}_{k\sigma})$ and $h_{k}(\tau)$ is the 
$2\times 2$ matrix defined as
\begin{eqnarray}
h_{k}(\tau)=
\left(\begin{array}{cc} \tilde{\epsilon}^c_{k} & -U^{\prime}\Delta^{\ast}_0+F(\tau)\\ 
-U^{\prime}\Delta_0+F(\tau) & \tilde{\epsilon}^f_{k}\\ \end{array} \right).
\label{eq:ham_mat}
\end{eqnarray}
In Eq. (\ref{eq:ham_mat}), we define 
$\tilde{\epsilon}^c_{k}=\epsilon^c_{k\sigma}+\frac{U}{2}n_c+U^{\prime}n_f$ and 
$\tilde{\epsilon}^f_{k}=\epsilon^f_{k\sigma}+\frac{U}{2}n_f+U^{\prime}n_c$ where 
$\epsilon^c_{k}=2t_c(\cos k_x+\cos k_y)+\mu_C$ and $\epsilon^f_{k}=2t_f(\cos k_x+\cos k_y)$ are the 
noninteracting energy dispersions for the $c$ and $f$ bands, respectively. 
In the ground state ($F_0=0$), $n_c$ and $\Delta_0$ are determined self-consistently. 

Within the Hartree-Fock approximation, the one-particle state at time 
$\tau$ with wave vector ${\bm k}$ and spin $\sigma$ is written as 
\begin{equation}
|\psi_{k\sigma}(\tau)\rangle = u_k(\tau)c^{\dagger}_{k\sigma}|0\rangle+v_k(\tau)f^{\dagger}_{k\sigma}|0\rangle, 
\end{equation}
with $|u_k(\tau)|^2+|v_k(\tau)|^2=1$. If $|\psi_{k\sigma}(\tau)\rangle$ is an occupied state at $\tau=0$, 
the momentum distribution function for the $c$ orbital,  
$n_c({\bm k})=\langle c^{\dagger}_{k\sigma}c_{k\sigma}\rangle$, and the electron-hole pair condensation 
in ${\bm k}$-space, $\Delta({\bm k})=\langle c^{\dagger}_{k\sigma}f_{k\sigma}\rangle$, are written as
\begin{equation}
n_c({\bm k})=|u_k(\tau)|^2,
\end{equation}
\begin{equation}
\Delta({\bm k})=u^{\ast}_k(\tau)v_k(\tau).
\end{equation}
In terms of these quantities, $n_c$ and $\Delta_0$ are given as
\begin{equation}
n_c=\frac{2}{N}\sum_{k}n_c({\bm k}),
\end{equation}
\begin{equation}
\Delta_0=\frac{1}{N}\sum_{k}\Delta({\bm k}). 
\end{equation}

The photoinduced dynamics is obtained by solving the time-dependent Schr$\ddot{\rm o}$dinger equation
\begin{equation}
|\psi_{k\sigma}(\tau+d\tau)\rangle = T\exp \Bigl[ -i\int^{\tau+d\tau}_{\tau}d\tau^{\prime}\hat{H}_{k\sigma}(\tau^{\prime})\Bigr]
|\psi_{k\sigma}(\tau)\rangle, 
\label{eq:schr_eq}
\end{equation}
where $T$ denotes the time-ordering operator. 
Equation (\ref{eq:schr_eq}) is numerically 
solved by writing\cite{Terai_TPS93,Kuwabara_JPSJ95,Tanaka_JPSJ10}
\begin{equation}
|\psi_{k\sigma}(\tau+d\tau)\rangle \simeq \exp \Bigl[ -id\tau \hat{H}_{k\sigma}(\tau+\frac{1}{2}d\tau)\Bigr]
|\psi_{k\sigma}(\tau)\rangle. 
\label{eq:schr_eq2}
\end{equation}
The exponential operator is expanded with time slice $d\tau=0.01$ until the norm of the wave 
function becomes unity with sufficient accuracy. 
For later convenience, we define the time average of a quantity $X(\tau)$ as
\begin{equation}
\overline{X}=\frac{1}{\tau_f-\tau_i}\int^{\tau_f}_{\tau_i}X(\tau)d\tau, 
\label{eq:time_av}
\end{equation}
where we use $\tau_i=80$ and $\tau_f=200$ throughout the study. 
We use $U=4$, $\mu_C=4$, and $\omega=0.4$ 
unless otherwise noted. The total number of sites is $N=800\times 800$. 

\section{Results}

\subsection{Ground-state properties}
\begin{figure}
\includegraphics[height=8.0cm]{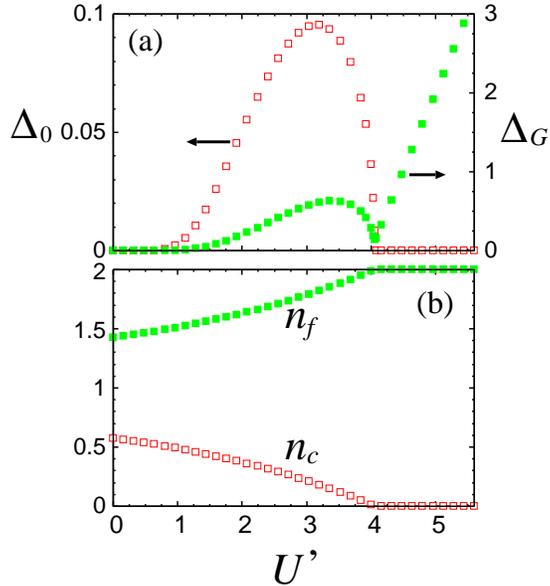}
\caption{(a) $\Delta_0$ and $\Delta_G$, (b) $n_c$ and $n_f$, as functions of $U^{\prime}$ with 
$U=4$, $\mu_C=4$.}
\label{fig:fig1}
\end{figure}

Before discussing photoinduced dynamics, we present ground-state properties. 
In Fig. \ref{fig:fig1}(a), we show the $U^{\prime}$ dependence of the excitonic order parameter 
$\Delta_0$ and the charge gap $\Delta_G$, where $\Delta_0$ is taken to be real. 
A similar plot for the electron densities $n_c$ and $n_f$ is shown in Fig. \ref{fig:fig1}(b). 
For $U^{\prime}=0$, the system is a metal with electron and hole pockets in the Fermi surface.  
Because of the perfect nesting of the Fermi surface, 
an infinitesimally small $U^{\prime}$ induces the excitonic 
order\cite{Watanabe_JPCS15,Kaneko_PRB12}, which results in $\Delta_G>0$. 
With increasing $U^{\prime}$, $\Delta_0$ exhibits a peak near $U^{\prime}=3.2$ 
and then decreases. At $U^{\prime}=U^{\prime}_{\rm cr}\sim 4.08$, 
a phase transition from the EI to a band insulator (BI) occurs. 
The density $n_c$ ($n_f$) monotonically decreases (increases) with increasing $U^{\prime}$ 
because of the Hartree shift\cite{Kaneko_PRB12}. 
In the BI phase, the $c$ and $f$ bands are completely decoupled so that we have 
$\Delta_0=0$ and $n_c=0$ ($n_f=2$). 
These results are qualitatively consistent with previous studies for ground states that take 
account of electron correlations\cite{Watanabe_JPCS15,Kaneko_PRB12}, where the properties of 
the excitonic order are discussed in the context of the 
BCS-BEC crossover\cite{Seki_PRB11,Phan_PRB10,Watanabe_JPCS15,Kaneko_PRB12}. When $U^{\prime}$ 
is small, the coherence length of electron-hole pairs $\xi$ is large and the order is well 
described by the BCS theory. On the other hand, in the region near the phase boundary between 
the EI and the BI, $\xi$ is small so that the BEC picture is more appropriate for describing 
the EI. 

\subsection{Photoinduced dynamics}

\begin{figure}
\includegraphics[height=8.0cm]{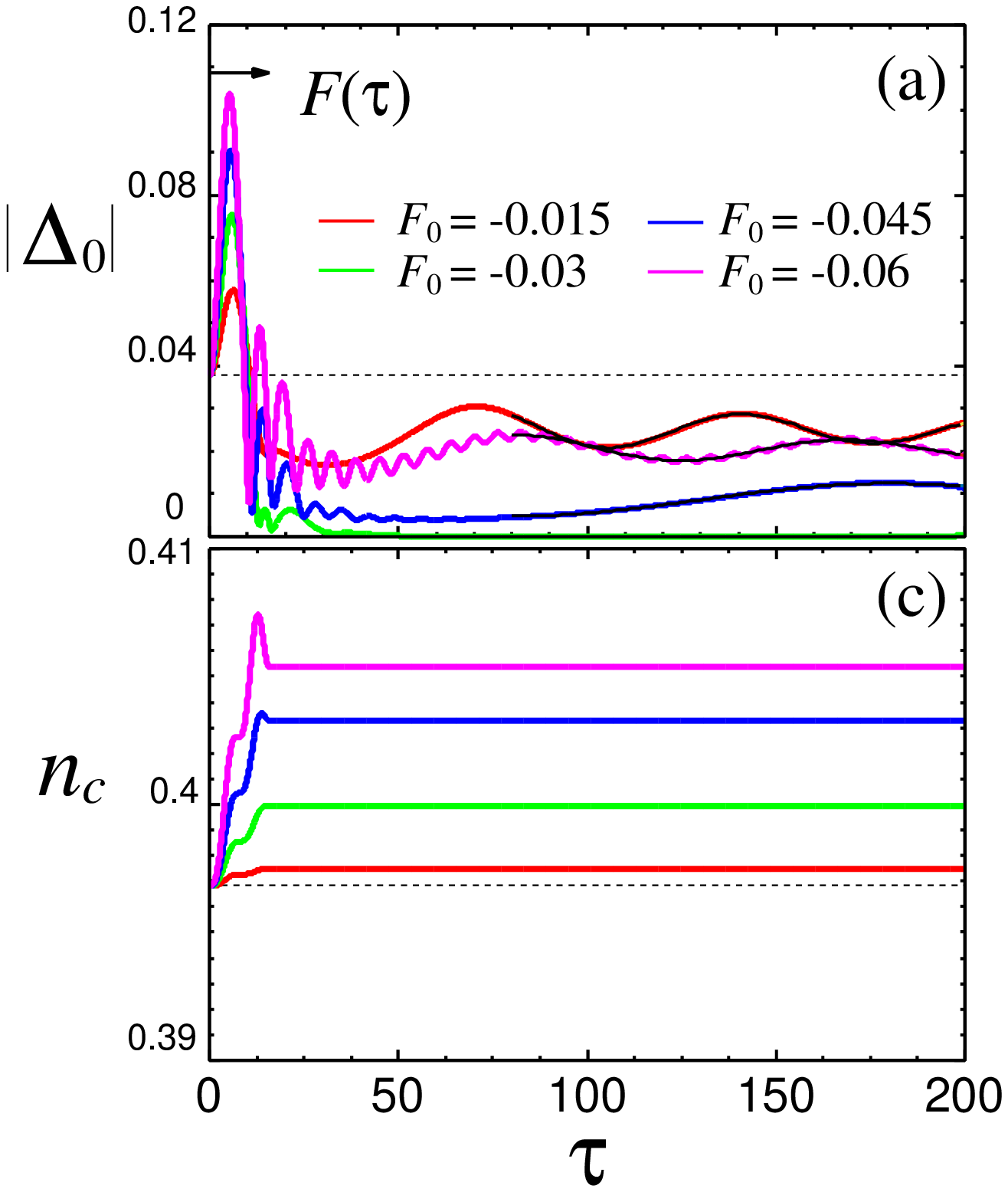}
\includegraphics[height=8.0cm]{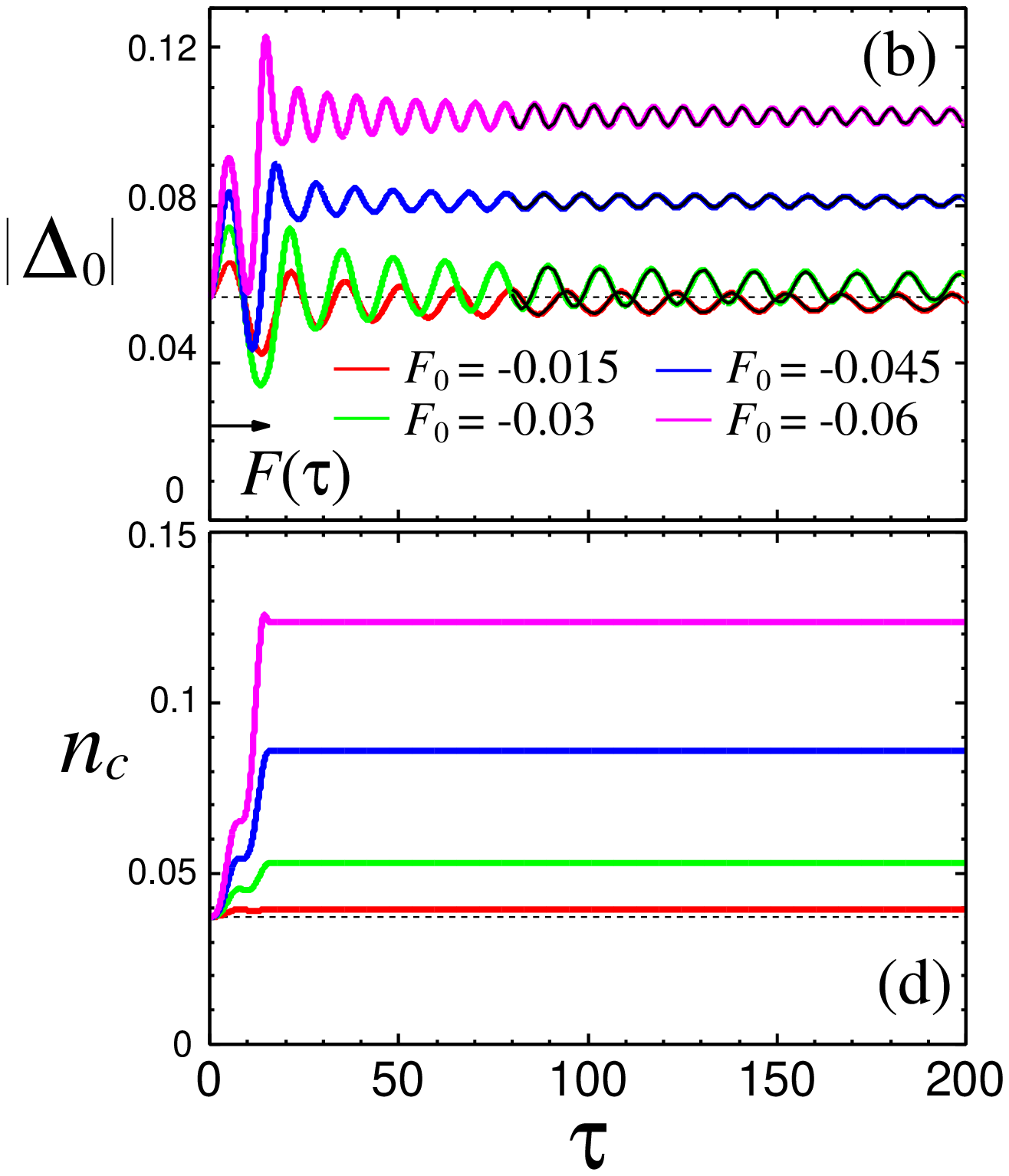}
\caption{Time evolution of $|\Delta_0|$ for (a) $U^{\prime}=1.8$ and (b) 
$U^{\prime}=3.9$ where the solid black lines 
are the fitting curves. The arrows indicate the range where $F(\tau)$ is nonzero. 
(c) and (d) show the time evolution of $n_c$ for $U^{\prime}=1.8$ and $3.9$, 
respectively. The horizontal dashed line in each panel indicates the corresponding 
equilibrium value. We use $U=4$, $\mu_C=4$, and $\omega=0.4$.}
\label{fig:fig2}
\end{figure}

\begin{figure}
\includegraphics[height=5.0cm]{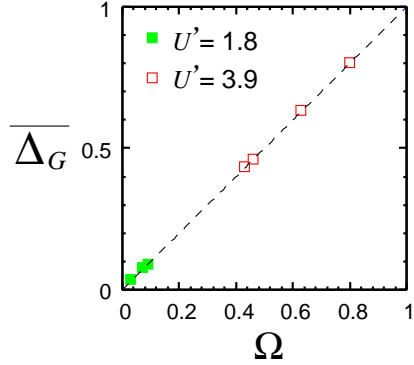}
\caption{Relation between $\Omega$ and time-averaged gap $\overline{\Delta_G}$. 
The solid (open) squares represent the results with $U^{\prime}=1.8$ ($U^{\prime}=3.9$). 
The dashed line indicates $\overline{\Delta_G}=\Omega$.}
\label{fig:fig3}
\end{figure}

In Figs. \ref{fig:fig2}(a) and \ref{fig:fig2}(b), we show the time evolution of $|\Delta_0|$ for 
$U^{\prime}=1.8$ and $3.9$ where the system is initially in the BCS and the BEC 
regimes, respectively. In Figs. 2(c) and 2(d), we depict the time profile of $n_c$. 
We use $F_0<0$, although our qualitative results are unaltered even if we use $F_0>0$. 
For $U^{\prime}=1.8$, $|\Delta_0|$ after the photoexcitation is 
smaller than that in the ground state, $|\Delta_0(\tau=0)|$. 
In particular, it almost vanishes for $F_0=-0.03$. 
On the other hand, for $U^{\prime}=3.9$, 
$|\Delta_0|$ becomes larger than $|\Delta_0(\tau=0)|$ as we increase $|F_0|$. 
After the photoexcitation, there is a characteristic oscillation in $|\Delta_0|$ for both 
$U^{\prime}=1.8$ and $U^{\prime}=3.9$. To analyze this oscillation, we fit a function, 
\begin{equation}
f(\tau)=A\tau^{-\gamma}\sin(\Omega \tau+\phi)+B, 
\end{equation}
to the time profile of $|\Delta_0|$ except for the 
cases where $|\Delta_0|\sim 0$ after the photoexcitation. We use the time range of 
$80<\tau<200$ to 
determine the parameters $A$, $B$, $\Omega$, $\gamma$, and $\phi$. The results are shown in 
Fig. \ref{fig:fig2}, indicating that they fit well to the curves. 
In Fig. \ref{fig:fig3}, we show the relation between $\Omega$ and $\overline{\Delta_G}$ that is 
the time average of the transient gap $\Delta_G(\tau)$. 
It is apparent that the frequency $\Omega$ corresponds to 
$\overline{\Delta_G}$, which demonstrates that the oscillation in $|\Delta_0|$ is the Higgs amplitude 
mode\cite{Volkov_JETP74,Littlewood_PRL81,Pekker_ARCMP15}. 
As shown in Figs. \ref{fig:fig2}(c) and \ref{fig:fig2}(d), $n_c$ is conserved 
for $F(\tau)=0$. In the case of $U^{\prime}=3.9$, the behavior of $n_c$ is similar to that of the 
time-averaged $|\Delta_0|$. When the time-averaged $|\Delta_0|$ is largely (slightly) enhanced after 
the photoexcitation, $n_c$ also increases largely (slightly) after that. 
In this case, the enhancement of $|\Delta_0|$ is interpreted as a consequence of the Hartree 
shift\cite{Murakami_arX17}. The increase in $n_c$ by the photoexcitation makes the two 
bands $\tilde{\epsilon}^c_k$ and $\tilde{\epsilon}^f_k$ approach each other, which promotes the mixing 
of these bands. However, for $U^{\prime}=1.8$, the behavior of $n_c$ is qualitatively 
different from that 
of $|\Delta_0|$. For example, the value of $|\Delta_0|$ after the 
photoexcitation with $F_0=-0.015$ and that with $F_0=-0.03$ are 
largely different although those of $n_c$ are slightly different. 
This indicates that the photoinduced change in $|\Delta_0|$ is not simply explained by the Hartree shift. 
To understand why $|\Delta_0|$ is enhanced (suppressed) in the BEC (BCS) regime 
more adequately, it is important to examine the momentum distribution function and the phase of the 
electron-hole pair condensation in ${\bm k}$-space, which will be discussed in Sect. III. C. 

\begin{figure}
\includegraphics[height=8.0cm]{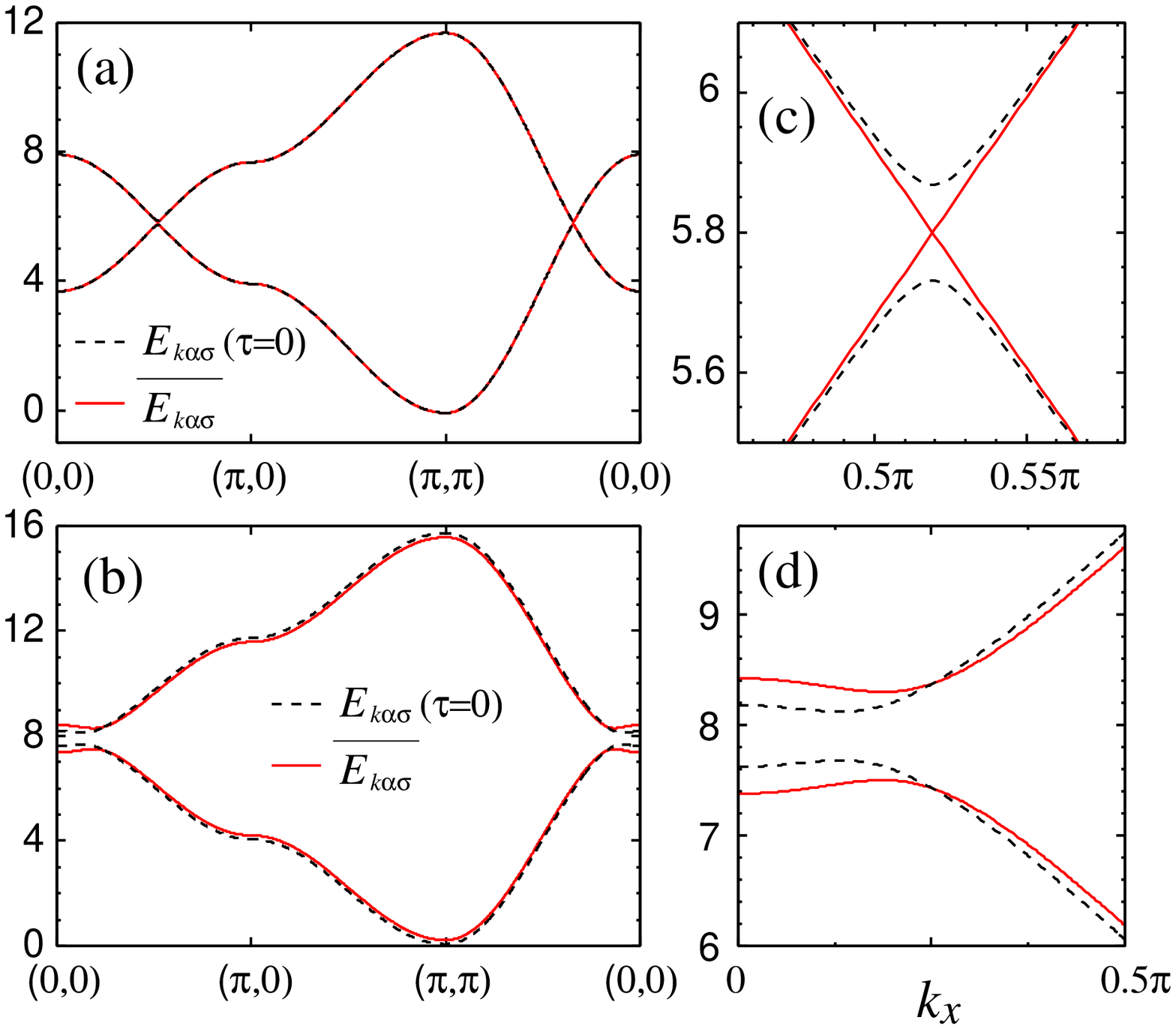}
\caption{Energy dispersion of bands in ground state $E_{k\alpha\sigma}(\tau=0)$ and 
time-averaged energy levels $\overline{E}_{k\alpha\sigma}$ for (a) $U^{\prime}=1.8$ 
and (b) $U^{\prime}=3.9$. In (a) [(b)], we show $\overline{E}_{k\alpha\sigma}$ with $F_0=-0.03$ 
($F_0=-0.06$). (c) [(d)] shows an enlarged view of (a) [(b)] near the initial gap located 
along the line between ${\bm k}=(0,0)$ and $(\pi,0)$.}
\label{fig:fig4}
\end{figure}

In Fig. \ref{fig:fig4}, we compare the energy dispersion of the bands in the ground state and 
the time-averaged energy levels after the photoexcitation. 
The latter is denoted by $\overline{E}_{k\alpha\sigma}$ with $\alpha$ being the band index. 
The quantity $\overline{E}_{k\alpha\sigma}$ is obtained from the time average of $E_{k\alpha\sigma}(\tau)$, 
which is the eigenvalue of $h_k(\tau)$. 
In Fig. \ref{fig:fig4}(a), we show the ground-state dispersion $E_{k\alpha\sigma}(\tau=0)$ 
and $\overline{E}_{k\alpha\sigma}$ with $F_0=-0.03$ for the case of 
$U^{\prime}=1.8$. An enlarged view near the initial gap is 
shown in Fig. \ref{fig:fig4}(c), indicating that the initial gap disappears in 
$\overline{E}_{k\alpha\sigma}$ since we have $|\Delta_0|\sim 0$ after 
the photoexcitation. Such a photoinduced gap closing has been reported recently 
in a system with an excitonic CDW by using the GW method that takes account of correlation effects 
beyond the mean-field theory\cite{Golez_PRB16}. Away from the gap, 
$E_{k\alpha\sigma}(\tau=0)$ and $\overline{E}_{k\alpha\sigma}$ 
are very close to each other because the change in $n_c$ by the 
photoexcitation is small as shown in Fig. \ref{fig:fig2}(c). 
Note that the difference between $E_{k\alpha\sigma}(\tau=0)$ and $\overline{E}_{k\alpha\sigma}$ 
near the gap originates from the change in $\Delta_0$, while that away from the gap comes 
from the change in $n_c$ ($n_f$). For the values of $F_0$ where 
$|\Delta_0|$ is nonzero after the photoexcitation (e.g. $F_0=-0.015$),  
the gap remains in $\overline{E}_{k\alpha\sigma}$ (not shown). 
For $U^{\prime}=3.9$, the charge gap is markedly enlarged because of the increase 
in $|\Delta_0|$. Away from ${\bm k}=(0,0)$, the upper and lower bands slightly approach to each other 
because the difference between the Hartree shifts for the two bands is reduced. 
We note that after the photoexcitation $E_{k\alpha\sigma}(\tau)$ has only a weak $\tau$ dependence so that 
$\overline{E}_{k\alpha\sigma}$ and $E_{k\alpha\sigma}(\tau)$ show similar ${\bm k}$ dependences even 
quantitatively. This is because $n_c$ and $n_f$ are conserved and the $\tau$ dependence comes only 
through $|\Delta_0|$ whose oscillation just affects energy levels in the vicinity 
of the gap. 

\begin{figure}
\includegraphics[height=4.0cm]{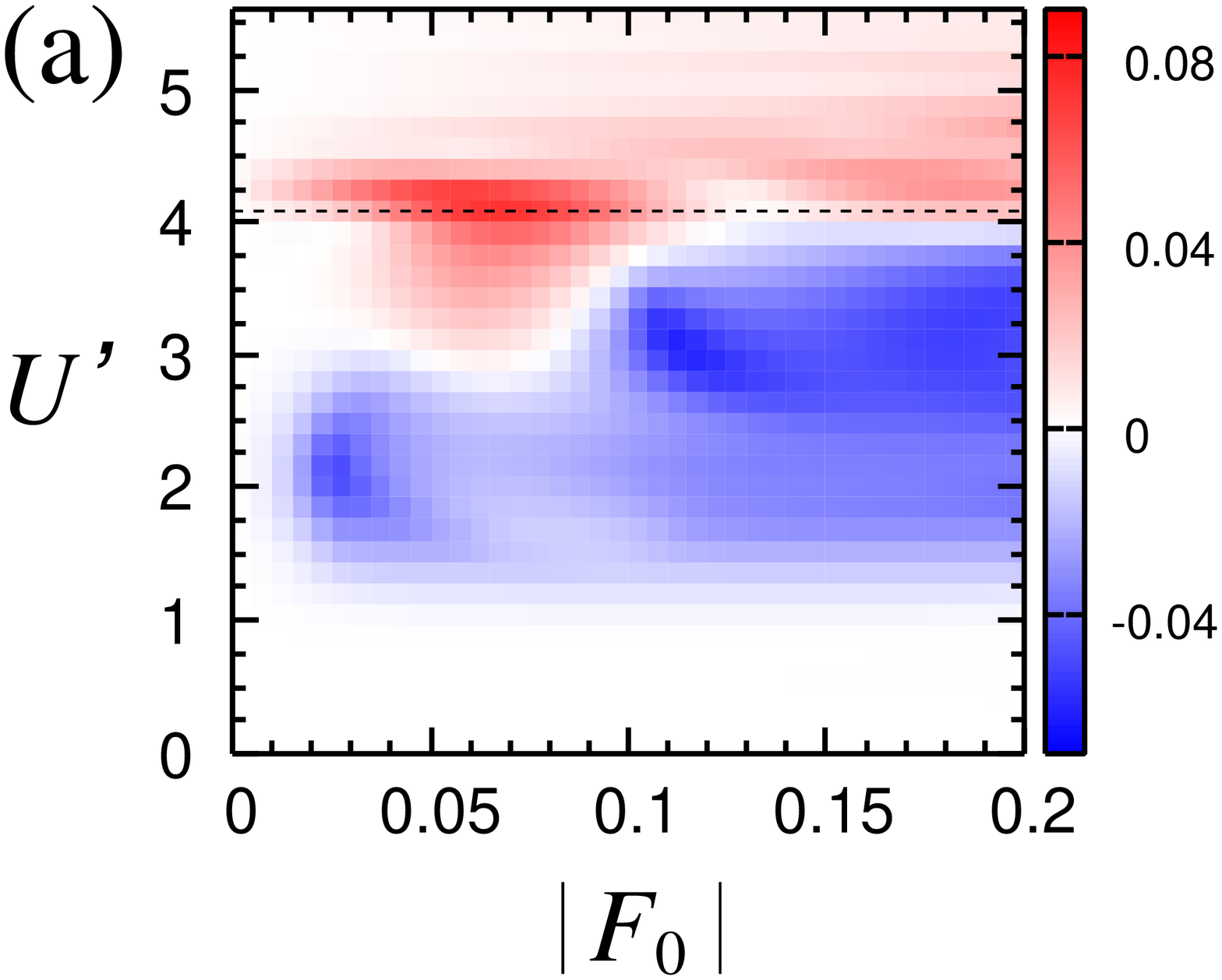}
\includegraphics[height=4.0cm]{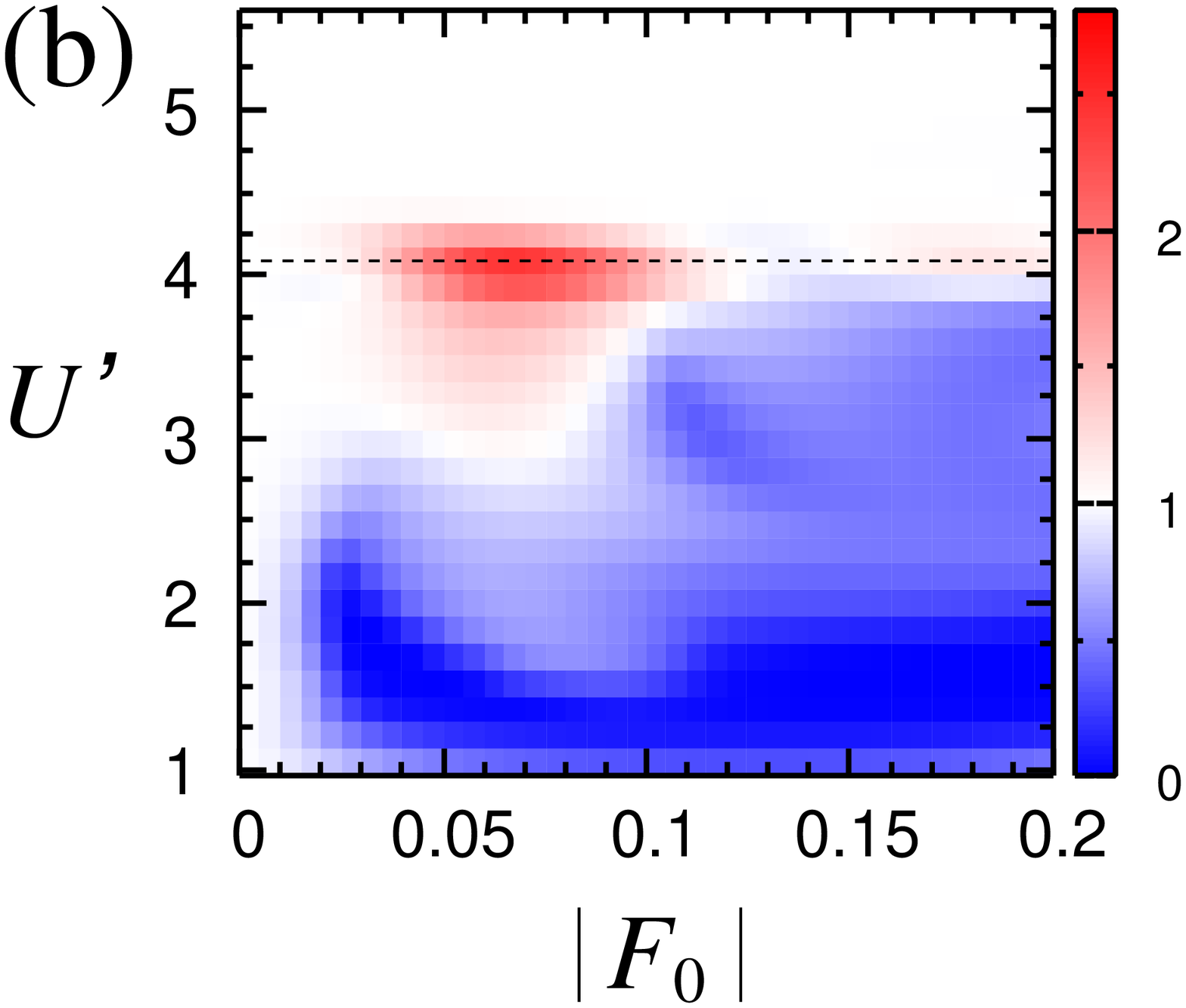}
\includegraphics[height=4.0cm]{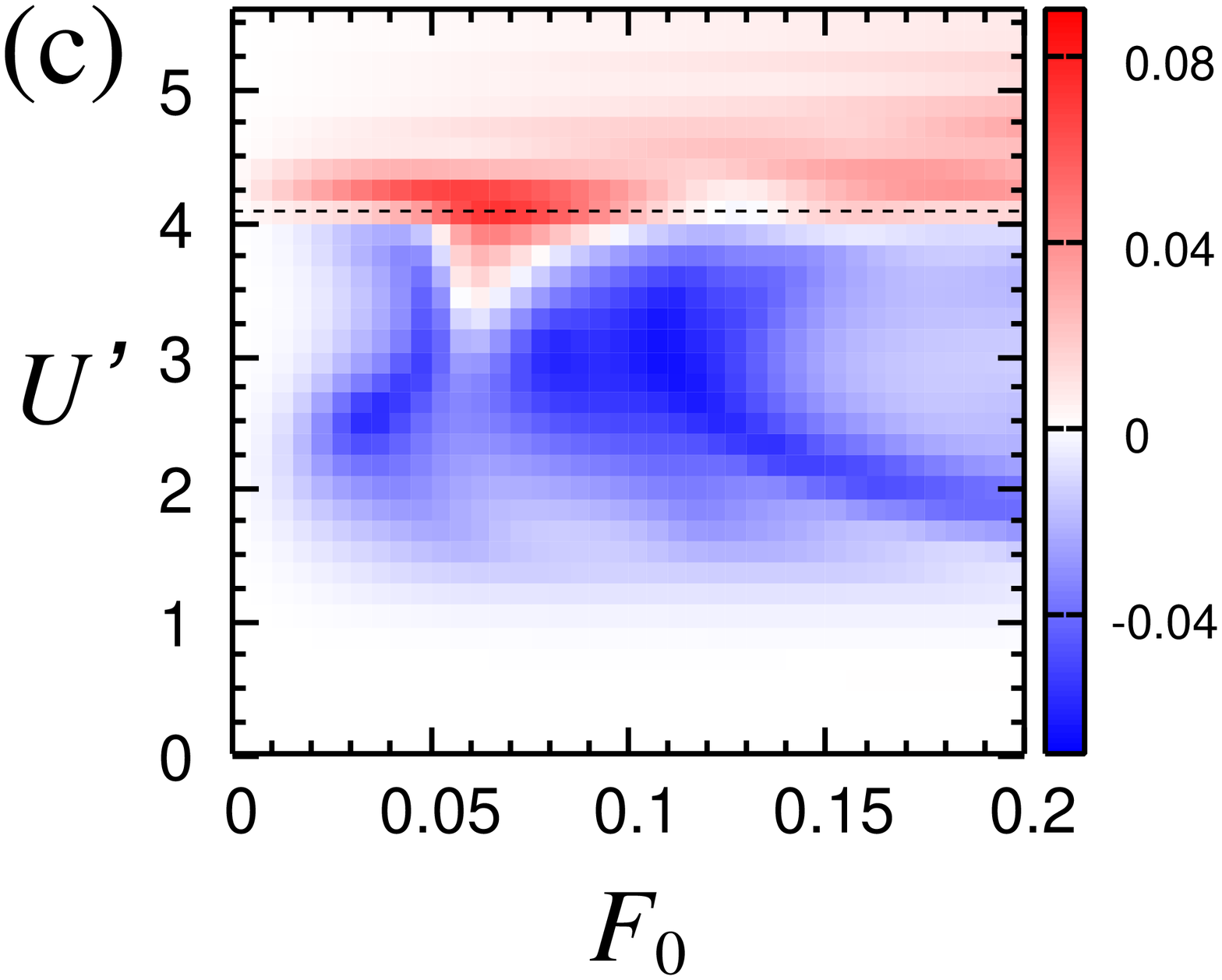}
\caption{(a) $\overline{|\Delta_{0}|}-\Delta_0(\tau=0)$ and 
(b) $\overline{\Delta_G}/\Delta_G(\tau=0)$ on $(|F_0|,U^{\prime})$ plane with $F_0<0$. 
(c) Similar plot for $\overline{|\Delta_{0}|}-\Delta_0(\tau=0)$ with $F_0>0$.
We use $U=4$, $\mu_C=4$, and $\omega=0.4$. 
The horizontal dashed lines indicate $U^{\prime}=U^{\prime}_{\rm cr}$.}
\label{fig:fig5}
\end{figure}

\begin{figure}
\includegraphics[height=4.0cm]{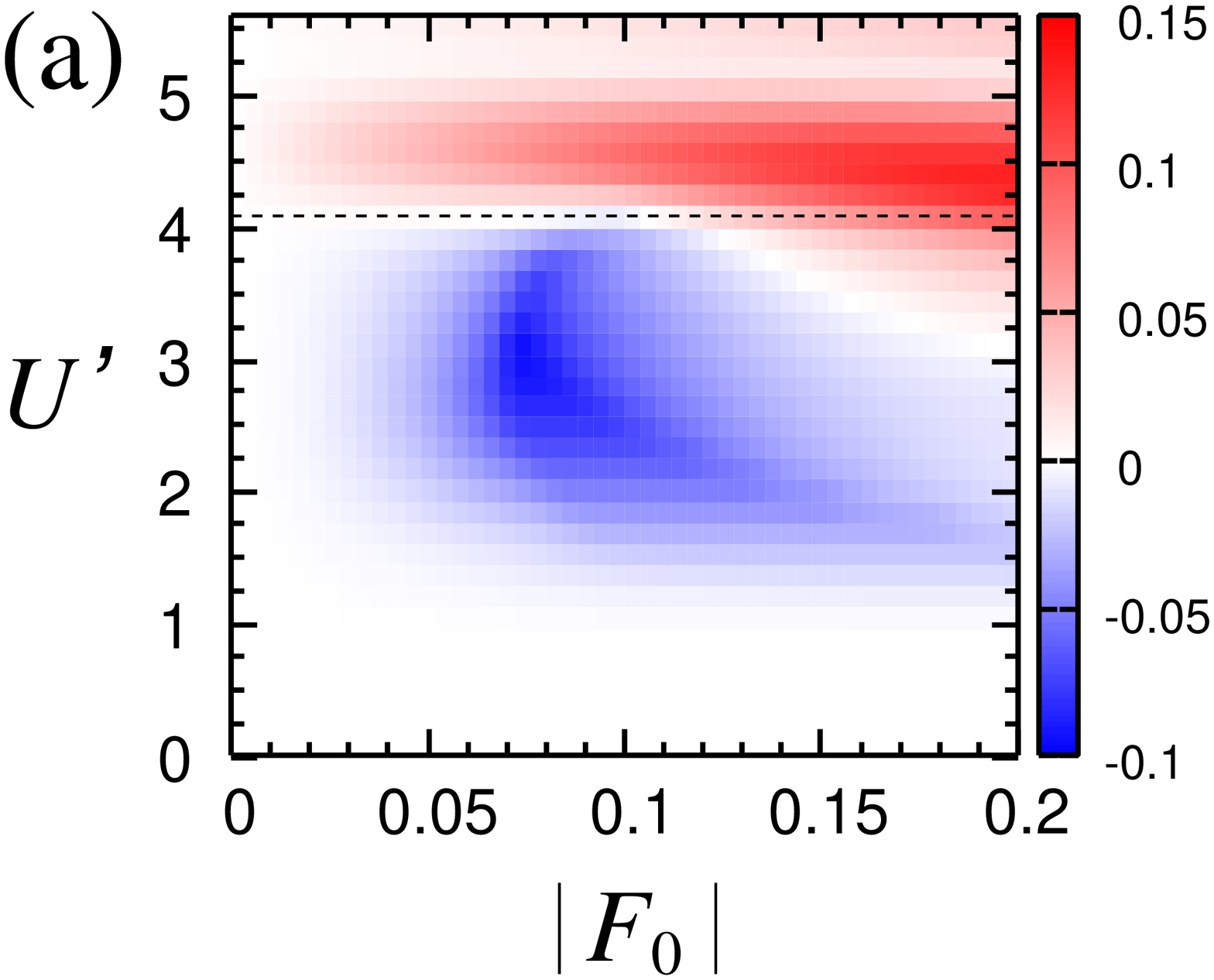}
\includegraphics[height=4.0cm]{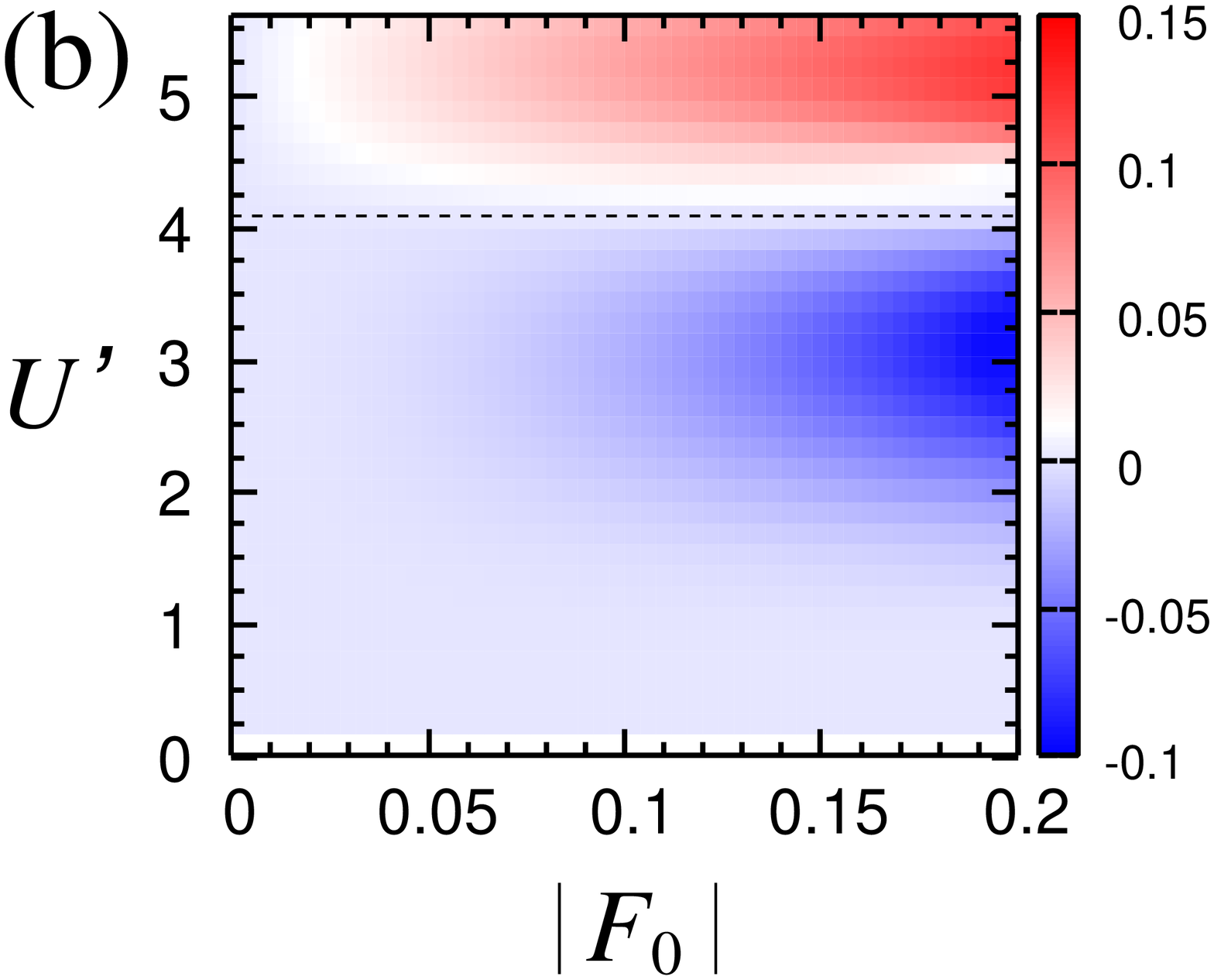}
\caption{$\overline{|\Delta_{0}|}-\Delta_0(\tau=0)$ on $(|F_0|,U^{\prime})$ plane 
with $F_0<0$ for (a) $\omega=1$ and (b) $\omega=2$.}
\label{fig:fig6}
\end{figure}

\begin{figure}
\includegraphics[height=5.0cm]{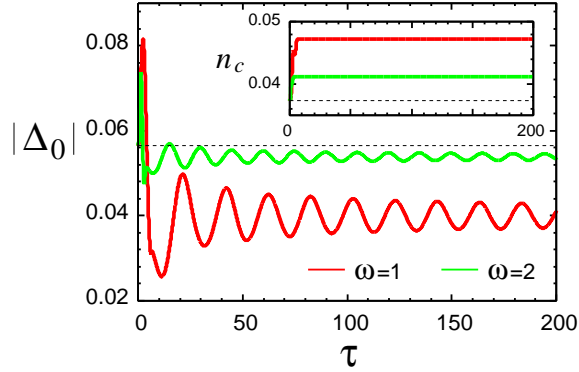}
\caption{Time evolution of $|\Delta_0|$ for $\omega=1$ and $\omega=2$. 
The inset shows the time evolution of $n_c$. We use $U^{\prime}=3.9$ and $F_0=-0.06$.}
\label{fig:fig7}
\end{figure}

Figure \ref{fig:fig5}(a) shows $\overline{|\Delta_{0}|}-\Delta_0(\tau=0)$ on the $(|F_0|,U^{\prime})$ plane. 
The excitonic order is largely enhanced around $U^{\prime}=U^{\prime}_{\rm cr}$. 
The enhancement occurs in a region where the initial state is in the BEC regime inside the 
EI phase and in the nearby BI phase where $\Delta_0(\tau=0)=0$. On the other hand, 
when $U^{\prime}$ is small and the system is initially in the BCS regime, 
$|\Delta_0|$ is decreased by photoexcitation. 
In Fig. \ref{fig:fig5}(b), we show $\overline{\Delta_G}/\Delta_G(\tau=0)$ 
whose enhancement is most prominent near $U^{\prime}=U^{\prime}_{\rm cr}$ and $|F_0|\sim 0.06$. 
For $U^{\prime}>U^{\prime}_{\rm cr}$, the enhancement is less clear compared to 
that of $\overline{|\Delta_{0}|}-\Delta_0(\tau=0)$. This is because the initial gap in the BI 
phase rapidly increases with $U^{\prime}$. 
In Fig. \ref{fig:fig5}(c), we show $\overline{|\Delta_{0}|}-\Delta_0(\tau=0)$ for the case 
of $F_0>0$, which indicates that the results are qualitatively the same as 
that for $F_0<0$. However, a quantitative difference appears depending on the sign of $F_0$, 
the reason of which will be discussed in Sect. III. C. 

In Fig. \ref{fig:fig6}, we show $\overline{|\Delta_{0}|}-\Delta_0(\tau=0)$ on the 
$(|F_0|,U^{\prime})$ plane with $F_0<0$ for $\omega=1$ and $\omega=2$. 
Compared to the case with $\omega=0.4$, the region where $\overline{|\Delta_{0}|}$ is enhanced 
shifts toward larger values 
of $|F_0|$ and $U^{\prime}$. In particular, there is almost no enhancement near 
$U^{\prime}=U^{\prime}_{\rm cr}$ for $\omega=2$.  
This comes from the fact that, when $\omega$ is much larger than the 
initial gap $\Delta_G(\tau=0)$, the charge transfer from the lower band to the upper band by the 
photoexcitation becomes ineffective so that the mixing of the two bands is hardly promoted. 
In Fig. \ref{fig:fig7}, we show the time profile of $|\Delta_0|$ and that of $n_c$ with 
$\omega=1$ and $\omega=2$ for $U^{\prime}=3.9$ where the initial 
gap is $\Delta_G(\tau=0)=0.44$. 
Changes in $|\Delta_0|$ and $n_c$ by the photoexcitation are small compared to 
those for $\omega=0.4$ shown in Figs. \ref{fig:fig2}(b) and \ref{fig:fig2}(d). 
$|\Delta_0|$ shows a characteristic oscillation with a frequency 
$\Omega=\overline{\Delta_G}$ corresponding to the Higgs amplitude mode as discussed above. 
After the photoexcitation, it becomes slightly smaller than $|\Delta_0(\tau=0)|$. 
Recently, Murakami {\it et al.} have shown that photoinduced enhancement of 
the excitonic order appears in a one-dimensional spinless fermion model with 
electron-phonon couplings\cite{Murakami_arX17}. They considered mainly the BEC-type excitonic 
order and used the external laser field with a frequency much larger than the initial gap. 
Without the electron-phonon couplings, they did not find any enhancement of the order, which is 
consistent with our results for large $\omega$. However, when the frequency $\omega$ is comparable 
to the initial gap, our results indicate that the excitonic order can be enhanced even in purely 
electronic systems, as shown in Fig. \ref{fig:fig5}. 

\subsection{Origin of photoinduced enhancement or suppression of $|\Delta_0|$} 

\begin{figure}
\includegraphics[height=8.0cm]{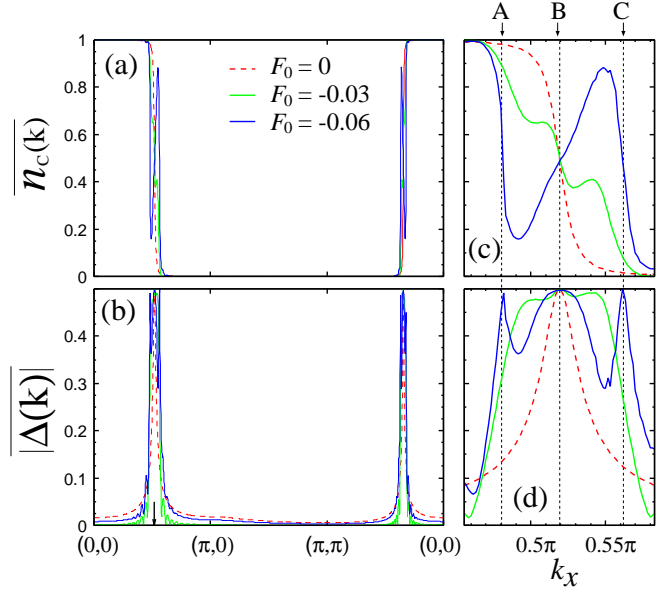}
\caption{(a) $\overline{n_c({\bm k})}$ and (b) $\overline{|\Delta({\bm k})|}$ for 
$U^{\prime}=1.8$. In (c) [(d)], an enlarged view of (a) [(b)] 
near ${\bm k}={\bm k}_F$ along the line from ${\bm k}=(0,0)$ to $(\pi,0)$ is shown. 
We use $F_0=0$, $-0.03$, and $-0.06$. The labels A, B, and C in (c) indicate the peak positions in 
$\overline{|\Delta({\bm k})|}$ for $F_0=-0.06$.}
\label{fig:fig8}
\end{figure}

\begin{figure}
\includegraphics[height=8.0cm]{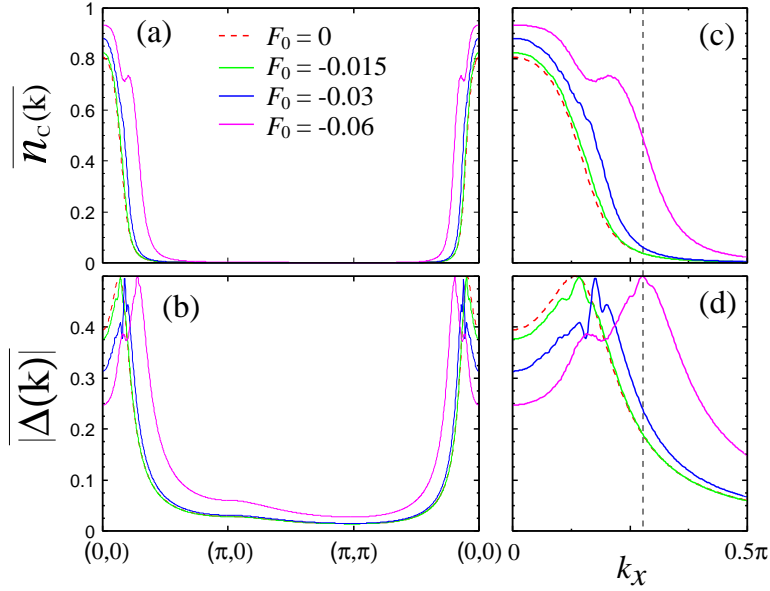}
\caption{(a) $\overline{n_c({\bm k})}$ and (b) $\overline{|\Delta({\bm k})|}$ for 
$U^{\prime}=3.9$. (c) [(d)] shows $\overline{n_c({\bm k})}$ ($\overline{|\Delta({\bm k})|}$) 
from ${\bm k}=(0,0)$ to $(\pi/2,0)$. We use $F_0=0$, $-0.015$, $-0.03$, and $-0.06$. 
The vertical line in (c) and (d) indicates the peak position in 
$\overline{|\Delta({\bm k})|}$ for $F_0=-0.06$.}
\label{fig:fig8b}
\end{figure}

Here, we discuss the origin of distinctive dynamics induced by the dipole transitions in 
the BCS and BEC regimes. 
In Figs. \ref{fig:fig8}(a) and \ref{fig:fig8}(b) [\ref{fig:fig8b}(a) and \ref{fig:fig8b}(b)], 
we show $\overline{n_c({\bm k})}$ and $\overline{|\Delta({\bm k})|}$, respectively, 
for $U^{\prime}=1.8$ ($U^{\prime}=3.9$) with different values of $F_0$. 
In the ground state, $n_c({\bm k})$ for $U^{\prime}=1.8$ exhibits a steep change along the lines 
from ${\bm k}=(0,0)$ to $(\pi,0)$ and from ${\bm k}=(\pi,\pi)$ to $(0,0)$, reflecting the 
energy dispersion shown in Fig. \ref{fig:fig4}(a). In Figs. \ref{fig:fig8}(c) and 
\ref{fig:fig8}(d), we show enlarged views of Figs. \ref{fig:fig8}(a) and \ref{fig:fig8}(b) near 
${\bm k}={\bm k}_F$, respectively, where we define   
${\bm k}_F=(k_{Fx},0)$ at which $n_c({\bm k}_F)=n_f({\bm k}_F)=0.5$ holds. At ${\bm k}={\bm k}_F$, 
$|\Delta({\bm k})|$ is sharply peaked at the maximum. 
We note that, because of Eqs. (7) and (8), $|\Delta({\bm k})|$ has its maximum value of 0.5 
when $n_c({\bm k})=n_f({\bm k})=0.5$. This relation holds even at $\tau>0$. 
The abrupt change in $n_c({\bm k})$ and $|\Delta({\bm k})|$ of the ground state in ${\bm k}$-space 
indicates the BCS nature of the EI. When $F_0$ is nonzero, $\overline{n_c({\bm k})}$ is strongly 
affected near ${\bm k}={\bm k_F}$. In particular, $\overline{n_c({\bm k})}$ decreases for $k_x<k_{Fx}$, 
whereas it increases for $k_x>k_{Fx}$. This means that, for $k_x<k_{Fx}$ ($k_x>k_{Fx}$), $c$-electrons 
($f$-electrons) are mainly transferred to the upper band by the photoexcitation. 
This characteristic ${\bm k}$ dependence 
of $\overline{n_c({\bm k})}$ cannot be explained by the Hartree shift. 
After the photoexcitation of $F_0=-0.06$, 
$\overline{|\Delta({\bm k})|}$ has three peaks near ${\bm k}={\bm k_F}$ 
which we label as A, B, and C in Fig. \ref{fig:fig8}(d). At these points, 
we have $\overline{n_c({\bm k})}\sim 0.5$ and $\overline{|\Delta({\bm k})|}\sim 0.5$. 
Both $\overline{n_c({\bm k})}$ 
and $\overline{|\Delta({\bm k})|}$ show abrupt changes in ${\bm k}$-space indicating 
that the excitonic order still has the BCS nature even after the photoexcitation. 
For $F_0=-0.03$ and $-0.06$, 
$\overline{|\Delta({\bm k})|}$ is enhanced around ${\bm k}={\bm k}_F$, however $\overline{|\Delta_0|}$ 
becomes smaller than $|\Delta_0(\tau=0)|$, as shown in Fig. \ref{fig:fig2}(a). This 
indicates that the photoinduced changes in the phases of $\Delta({\bm k})$ depend strongly on ${\bm k}$ 
in the Brillouin zone. 
For $U^{\prime}=3.9$, 
$n_c({\bm k})$ and $|\Delta({\bm k})|$ of the ground state gradually vary with ${\bm k}$, 
as shown in Figs. \ref{fig:fig8b}(a) and \ref{fig:fig8b}(b), 
respectively, because of the BEC nature of the 
EI. In contrast to the case of $U^{\prime}=1.8$, $\overline{n_c({\bm k})}$ is increased 
by the photoexcitation for all ${\bm k}$. In Figs. \ref{fig:fig8b}(c) and \ref{fig:fig8b}(d), 
we show $\overline{n_c({\bm k})}$ and $\overline{|\Delta({\bm k})|}$ from ${\bm k}=(0,0)$ to $(\pi/2,0)$. 
Although 
$\overline{|\Delta({\bm k})|}$ decreases around ${\bm k}=(0,0)$, it increases in a large area of the Brillouin 
zone. The BEC nature of the excitonic order is maintained through the photoexcitation since 
$\overline{n_c({\bm k})}$ and $\overline{|\Delta({\bm k})|}$ gradually change with ${\bm k}$ as in the ground state. 

\begin{figure}
\includegraphics[height=4.0cm]{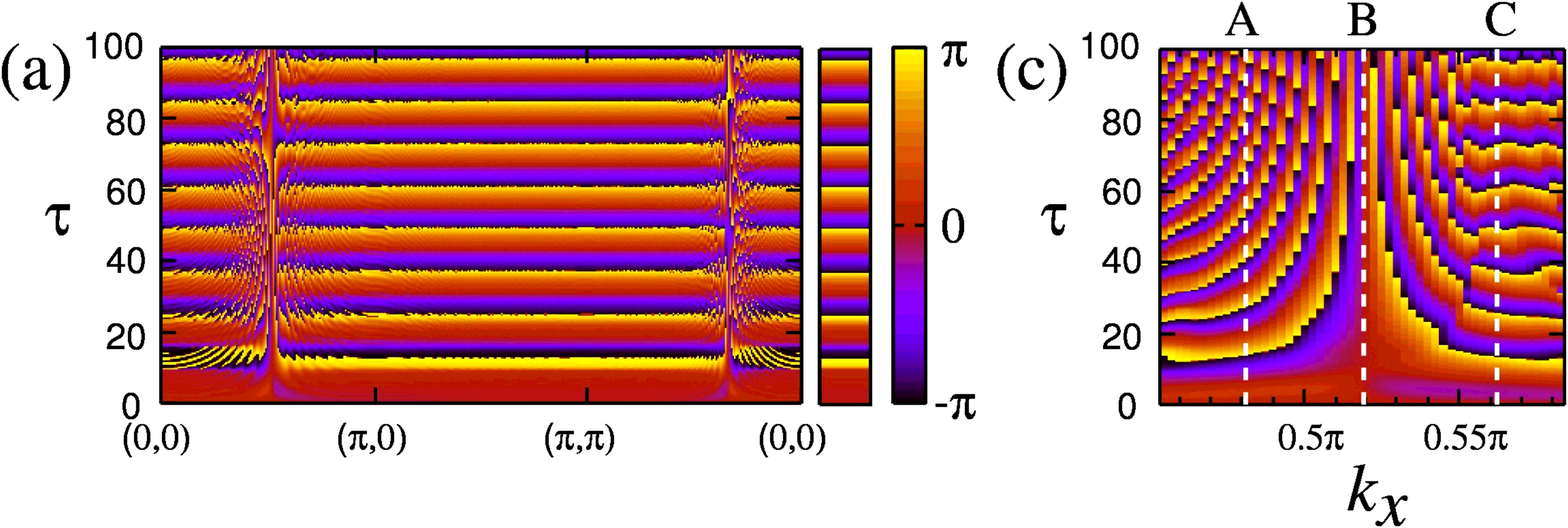}
\includegraphics[height=4.0cm]{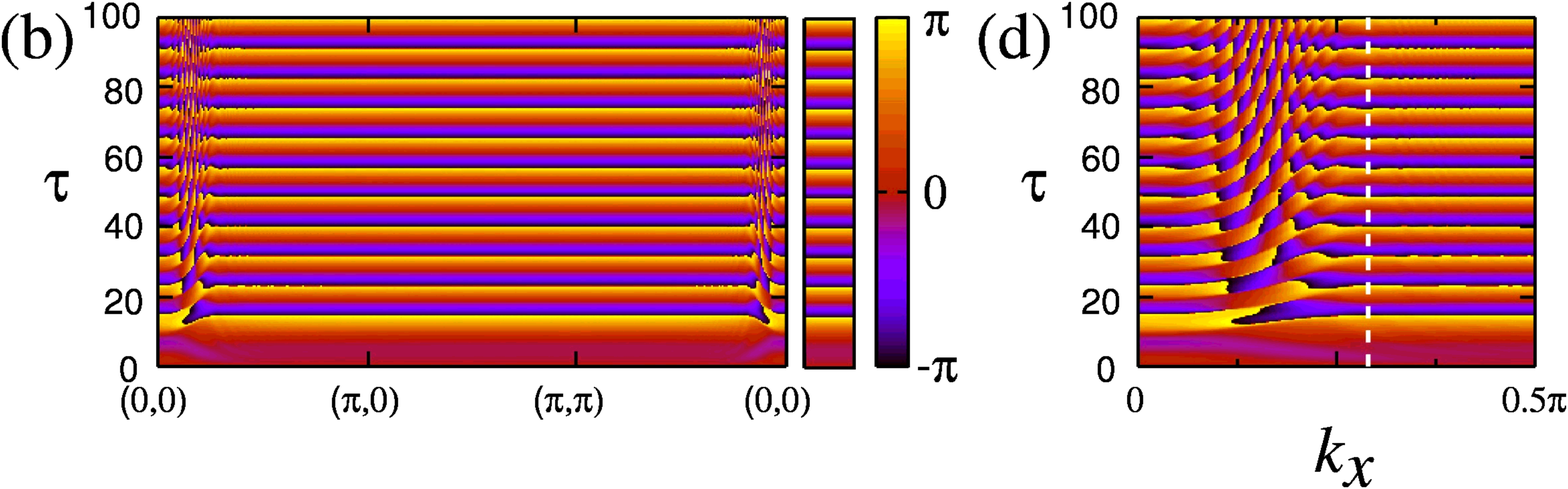}
\caption{Time evolution of $\theta_k$ for (a) $U^{\prime}=1.8$ and 
(b) $U^{\prime}=3.9$ with $F_0=-0.06$.  
The right panel shows the time evolution of $\theta$. (c) [(d)] shows an 
enlarged view of (a) [(b)] as in Fig. \ref{fig:fig4}. In (d), the dotted line indicates the 
peak position of $\overline{|\Delta({\bm k})|}$ in Fig. \ref{fig:fig8b}(d).}
\label{fig:fig9}
\end{figure}

\begin{figure}
\includegraphics[height=4.0cm]{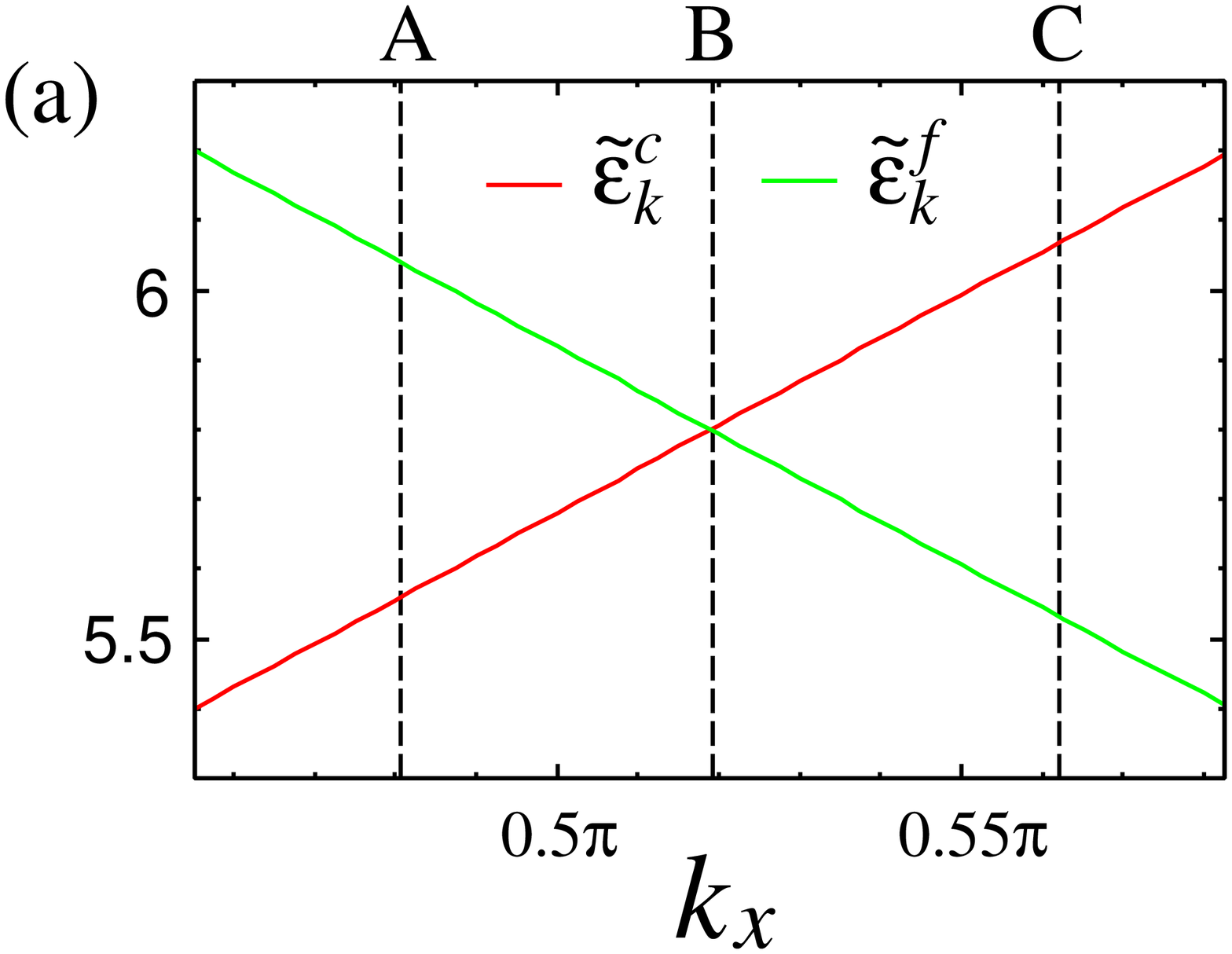}
\includegraphics[height=4.0cm]{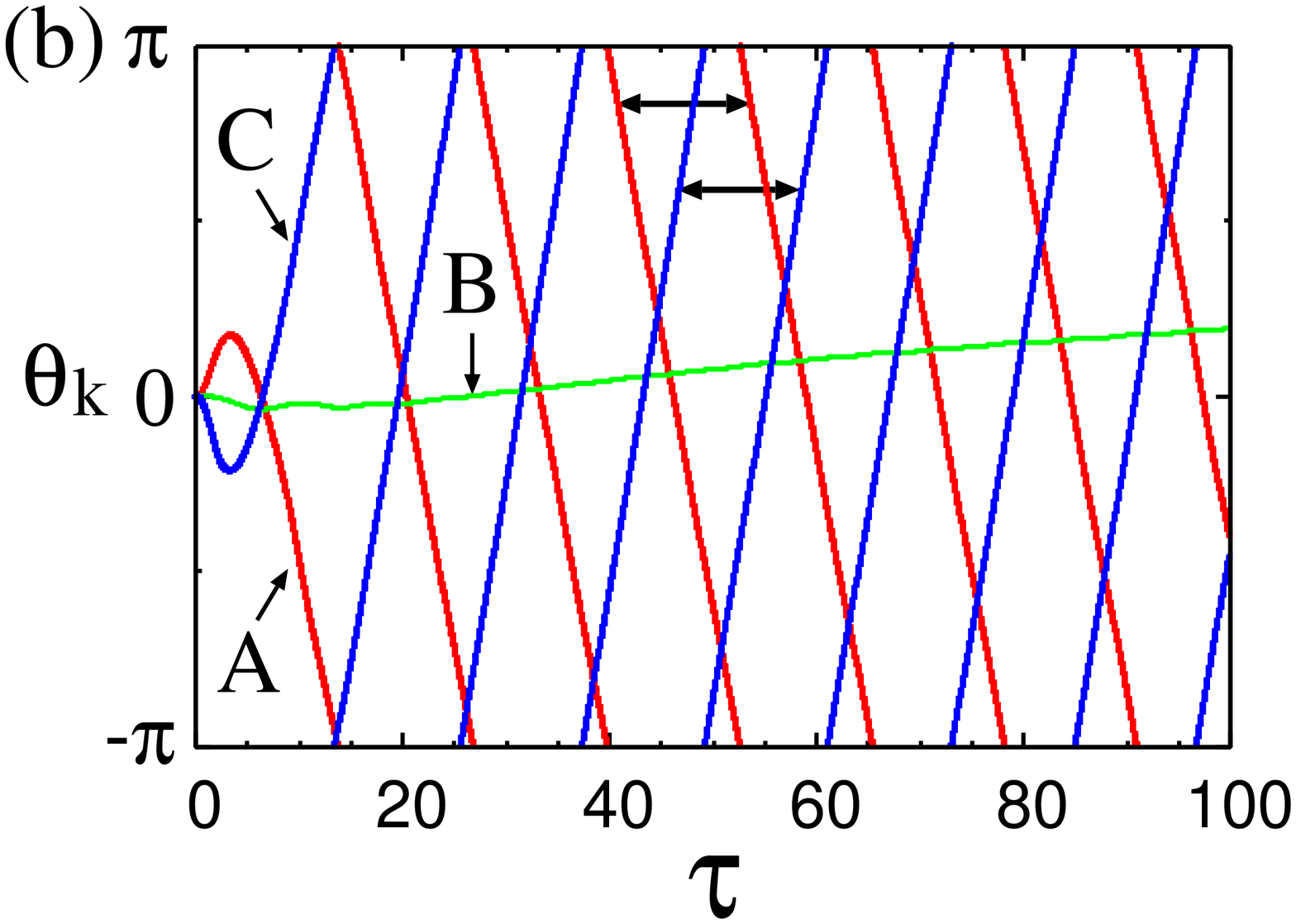}
\caption{(a) $\tilde{\epsilon}^c_k$ and $\tilde{\epsilon}^f_k$ near ${\bm k}={\bm k}_F$ 
after photoexcitation. (b) Time evolution of $\theta_k$ at A, B, and C. We use 
$U^{\prime}=1.8$ and $F_0=-0.06$. The double-headed 
arrows indicate $T_k$ estimated from $\tilde{\epsilon}^c_k$ and $\tilde{\epsilon}^f_k$ at A and C.} 
\label{fig:fig10}
\end{figure}

In Figs. \ref{fig:fig9}(a) and \ref{fig:fig9}(b), 
we show the time profile of the phase of $\Delta({\bm k})=|\Delta({\bm k})|e^{i\theta_k}$ 
for $U^{\prime}=1.8$ and $U^{\prime}=3.9$ in the case of $F_0=-0.06$. 
For comparison, we also show the phase $\theta$ of $\Delta_0$ ($=|\Delta_0| e^{i\theta}$) 
in the right panels. We recall that $\Delta_0$ and $\Delta({\bm k})$ are related by Eq. (10). 
In Figs. \ref{fig:fig9}(c) and \ref{fig:fig9}(d), we show an enlarged view of 
Figs. \ref{fig:fig9}(a) and \ref{fig:fig9}(b), respectively, as in Fig. \ref{fig:fig4}. 
For $U^{\prime}=1.8$, the time profile of $\theta_k$ depends strongly on ${\bm k}$ in the region 
where $\overline{n_c({\bm k})}$ and $\overline{|\Delta({\bm k})|}$ show abrupt changes. 
Away from this region, their time profiles become in phase with that of $\theta$. 
On the other hand, for $U^{\prime}=3.9$, 
$\theta_k$ shows only a weak ${\bm k}$ dependence except for the region near the $\Gamma$ point, and their 
time profiles are in phase with that of $\theta$ in a wide area of the Brillouin zone. In particular, 
$\theta_k$ at the peak position of $\overline{|\Delta({\bm k})|}$ is in phase with that of $\theta$ as 
shown in Fig. \ref{fig:fig9}(d), which is in sharp contrast to the case of $U^{\prime}=1.8$. 
The behavior of $\theta_k$ for $U^{\prime}=1.8$ 
is analyzed by using the equation of motion for $\Delta({\bm k})$ written as
\begin{eqnarray}
\partial_{\tau} \Delta({\bm k})&=&-i\langle [ c^{\dagger}_{k\sigma}f_{k\sigma},\hat{H}^{\rm HF}_{\rm tot}(\tau)]\rangle \\ \nonumber
&=& i(\tilde{\epsilon}^c_k-\tilde{\epsilon}^f_k)\Delta({\bm k}) \\ \nonumber
&+&i(-U^{\prime}\Delta_0+F(\tau))(n_f({\bm k})-n_c({\bm k})).
\end{eqnarray}
After the photoexcitation, $F(\tau)=0$ and the second term on the right-hand side of Eq. (15) 
becomes very small at A, B, and C in Fig. \ref{fig:fig8}(c) because of 
the relation $n_c({\bm k})\sim n_f({\bm k})\sim 0.5$. Therefore, at these ${\bm k}$ points, 
$\tilde{\epsilon}^c_k-\tilde{\epsilon}^f_k$ essentially determines the time evolution of $\theta_k$ 
and the period of $\theta_k$ should be given by $T_k=2\pi/|\tilde{\epsilon}^c_k-\tilde{\epsilon}^f_k|$. 
In Fig. \ref{fig:fig10}(a), we show $\tilde{\epsilon}^c_k$ and $\tilde{\epsilon}^f_k$ near 
${\bm k}={\bm k}_F$ after the 
photoexcitation, whereas the time evolution of $\theta_k$ at A, B, and C is depicted 
in Fig. \ref{fig:fig10}(b). We have $\dot{\theta}_k<0$, $\dot{\theta}_k\sim 0$, and $\dot{\theta}_k>0$ at 
A, B, and C, reflecting $\tilde{\epsilon}^c_k<\tilde{\epsilon}^f_k$, 
$\tilde{\epsilon}^c_k\sim \tilde{\epsilon}^f_k$, and 
$\tilde{\epsilon}^c_k>\tilde{\epsilon}^f_k$ as shown in Fig. \ref{fig:fig10}(a), respectively. 
Moreover, at A and C, the period of $\theta_k$ coincides with $T_k$ estimated from 
$\tilde{\epsilon}^c_k$ and $\tilde{\epsilon}^f_k$, 
demonstrating that the results are consistent with the above arguments. 
These arguments show that for the ${\bm k}$ region where $|\Delta({\bm k})|$ is large, the time profiles 
of the phases of 
$\Delta({\bm k})$ are out of phase for $U^{\prime}=1.8$, which inhibits the enhancement of $|\Delta_0|$. 
For $U^{\prime}=3.9$, since the time profile of $\theta_k$ is in phase with that of $\theta$, 
the increase in $|\Delta({\bm k})|$ in the large area of the Brillouin zone gives rise to the enhancement of 
$|\Delta_0|$. 
In this case, the dynamics of the order parameter is well described by the real space picture. 

\begin{figure}
\includegraphics[height=8.0cm]{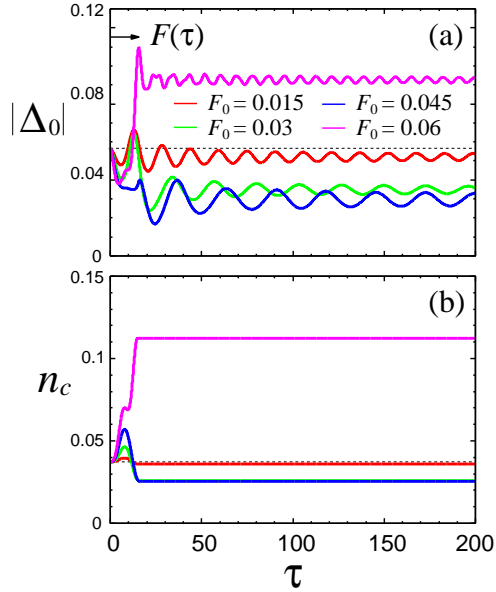}
\caption{A similar plot with Fig. \ref{fig:fig2} for the case of $F_0>0$. 
We use $U^{\prime}=3.9$ and $\omega=0.4$.}
\label{fig:fig_FP}
\end{figure}

With these results, we discuss the detailed structure of Fig. \ref{fig:fig5}. 
When $U^{\prime}$ is small and the initial EI state is in the BCS regime, 
$|\Delta_0|$ is basically suppressed by the photoexcitation regardless of the values of $F_0$. 
This is because the time profile of $\theta_k$ strongly depends on ${\bm k}$ especially near 
the peak positions of 
$\overline{|\Delta({\bm k})|}$ as shown in Figs. \ref{fig:fig9}(a) and \ref{fig:fig9}(c). 
When $U^{\prime}$ is large and the initial EI state is in the BEC regime, $|\Delta_0|$ is 
slightly reduced by the photoexcitation for small $|F_0|$. 
In order to explain the reason 
for this, we first discuss the effect of the sign of $F_0$ on our results. 
At $\tau=0$, we choose $\Delta_0$ as real and positive. In this case, the 
magnitude of the off-diagonal elements of $h_k(\tau)$ initially increases if $F_0<0$, 
whereas it decreases if $F_0>0$. This initial increase (decrease) results in an 
enhancement (suppression) of $|\Delta_0|$ just after the external field is switched on, as 
observed for the case of $F_0<0$ in Figs. \ref{fig:fig2}(a) and \ref{fig:fig2}(b). 
We show the time profile of 
$|\Delta_0|$ and $n_c$ for $F_0>0$ with $U^{\prime}=3.9$ in Fig. \ref{fig:fig_FP}, which in fact 
indicates that $|\Delta_0|$ initially decreases when $F(\tau)$ is switched on. By this effect, 
the region where $|\Delta_0|$ is enhanced for $F_0>0$ is narrower than that for $F_0<0$ as shown in 
Figs. \ref{fig:fig5}(a) and \ref{fig:fig5}(c). For small $|F_0|$ with $F_0>0$ ($F_0<0.05$), the initial 
reduction of $|\Delta_0|$ dominates the behavior of $|\Delta_0|$. For $F_0\sim 0.06$, 
the charge transfer from the lower band to the upper band dominates over this 
effect, so that $|\Delta_0|$ is enhanced even in the case of $F_0>0$. 
For small $|F_0|$ ($|F_0|<0.02$) with $F_0<0$, $|\Delta_0|$ is slightly suppressed for large 
$U^{\prime}$ since $\overline{|\Delta({\bm k})|}$ becomes smaller than that in the ground state around 
the $\Gamma$ point as shown in Fig. \ref{fig:fig8b}(d). We note that the time profile of $\theta_k$ 
slightly deviates from that of $\theta$ in this region, whereas it is almost in phase with that of 
$\theta$ away from the $\Gamma$ point. 

\begin{figure}
\includegraphics[height=8.0cm]{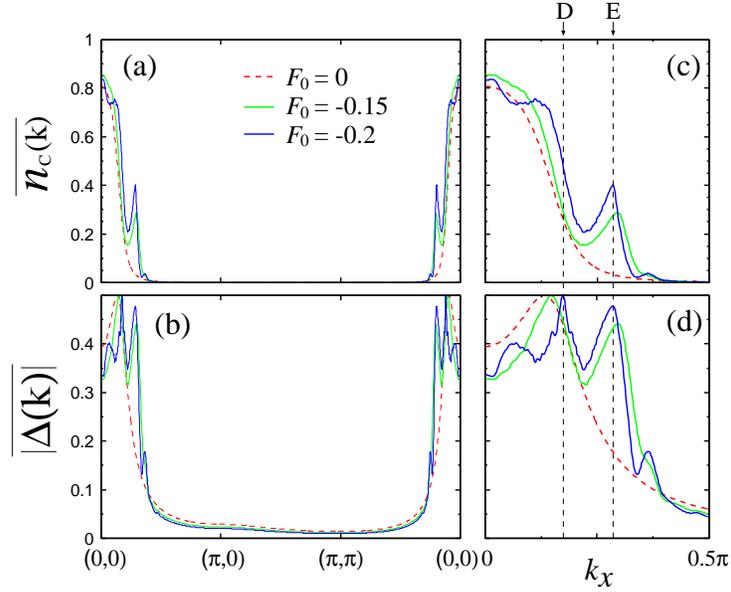}
\caption{(a) $\overline{n_c({\bm k})}$ and (b) $\overline{|\Delta({\bm k})|}$ for 
$U^{\prime}=3.9$. We use $F_0=0$, $-0.15$, and $-0.2$.  
(c) [(d)] shows $\overline{n_c({\bm k})}$ ($\overline{|\Delta({\bm k})|}$) 
from ${\bm k}=(0,0)$ to $(\pi/2,0)$.}
\label{fig:fig11}
\end{figure}

\begin{figure}
\includegraphics[height=4.0cm]{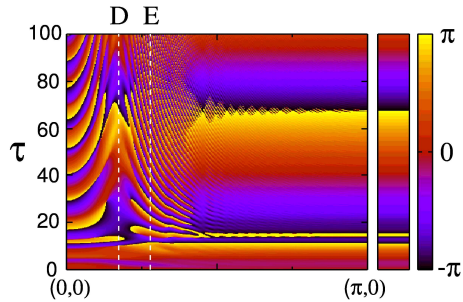}
\caption{Time evolution of $\theta_k$ for $U^{\prime}=3.9$ with $F_0=-0.2$ 
from ${\bm k}=(0,0)$ to $(\pi,0)$. The right panel shows the time evolution of $\theta$.}
\label{fig:fig12}
\end{figure}

When $U^{\prime}$ is slightly smaller than $U^{\prime}_{\rm cr}$, $|\Delta_0|$ 
is suppressed compared to $\Delta_0(\tau=0)$ in a region of large $|F_0|$ ($|F_0|>0.12$), 
although it is enhanced around $|F_0|\sim 0.06$. 
In Fig. \ref{fig:fig11}(a) [\ref{fig:fig11}(b)], we show 
$\overline{n_c({\bm k})}$ 
($\overline{|\Delta({\bm k})|}$) for $F_0=-0.15$ and $-0.2$ where we have 
$\overline{|\Delta_0|}<\Delta_0(\tau=0)$. Between ${\bm k}=(0,0)$ and $(\pi,0)$, $\overline{|\Delta({\bm k})|}$ 
has two large peaks as shown in Fig. \ref{fig:fig11}(d), which we label as D and E for the case of $F_0=-0.2$. 
The peak at D appears since $\overline{n_c({\bm k})}$ crosses 0.5, whereas E reflects a sharp peak in 
$\overline{n_c({\bm k})}$. Although $\overline{|\Delta({\bm k})|}$ is enhanced near the point E, their values 
away from the two peaks are slightly smaller than those in the ground state. In Fig. \ref{fig:fig12}, we show 
the time profile of $\theta_k$ from ${\bm k}=(0,0)$ to $(\pi,0)$ for $F_0=-0.2$. 
Notably, $\theta_k$ strongly depends on ${\bm k}$ in a region including D and E. 
The time profile of $\theta_k$ at E is quite different from that of $\theta$. 
Although the time profile 
of $\theta_k$ away from this region is in phase with that of $\theta$, the values of 
$\overline{|\Delta({\bm k})|}$ are smaller than those in the ground state, so that they do not 
contribute to enhance $|\Delta_0|$. These results indicate 
that even when the initial state is the BEC-type EI, whether $|\Delta_0|$ is enhanced or not depends on the 
value of $F_0$.

\section{Summary}
In this paper, we study dipole-transition-induced dynamics of excitonic orders by using the two-orbital 
Hubbard model on the square lattice. We show that the photoinduced dynamics depends strongly on whether 
the EI is initially in the BCS regime or in the BEC regime. The excitonic order is basically enhanced 
by the photoexcitation in the latter, whereas it is reduced in the former. These results are caused by different 
behaviors of the momentum distribution functions and the phases of the electron-hole pair condensation 
in ${\bm k}$-space. When the initial EI is of the BEC-type, its dynamics is interpreted from the real 
space picture, whereas the ${\bm k}$ dependence of physical quantities is essential for the BCS-type EI. 

\begin{acknowledgments}
The authors thank Y. Murakami for fruitful discussions. This work was supported by Grants-in-Aid 
for Scientific Research (C) (Grant No. 16K05459) and Scientific Research (A) (Grant No. 15H02100) 
from the Ministry of Education, Culture, Sports, Science and Technology of Japan. 
\end{acknowledgments}

\end{document}